\documentclass[11pt]{article}
\usepackage{etoolbox,xparse}
\usepackage{amsmath, amssymb, amsthm}
\usepackage{fullpage}
\usepackage{setspace,color,physics}
\usepackage{url,hyperref}
\usepackage[shortlabels]{enumitem}
\usepackage{tikz}
\usetikzlibrary{positioning}

\newtheorem{theorem}{Theorem}[section]
\newtheorem*{theorem*}{Theorem}
\newtheorem{cor}[theorem]{Corollary}
\newtheorem{prop}[theorem]{Proposition}
\newtheorem{lemma}[theorem]{Lemma}

\theoremstyle{definition}
\newtheorem{define}{Definition}[section]
\newtheorem*{define*}{Definition}

\theoremstyle{remark}
\newtheorem*{remark}{Remark}

\let\q\qty

\let\phi\varphi

\let\epsilon\varepsilon

\newcommand{\set}[1]{\left\{#1\right\}}

\newcommand{\without}[1]{\setminus \set{#1}}

\newcommand{\qlfloor}{\left\lfloor}
\newcommand{\qrfloor}{\right\rfloor}
\newcommand{\qlceil}{\left\lceil}
\newcommand{\qrceil}{\right\rceil}
\newcommand{\floor}[1]{\qlfloor #1 \qrfloor}
\newcommand{\ceil}[1]{\qlceil #1 \qrceil}

\DeclareMathOperator{\polylog}{polylog}
\newcommand{\mrm}{\mathrm}
\newcommand{\msf}{\mathsf}

\newcommand{\TIME}[1][{}]{\mathsf{TIME}\left[#1\right]}
\newcommand{\SPACE}[1][{}]{\mathsf{SPACE}\left[#1\right]}
\renewcommand{\L}{\msf{L}}
\renewcommand{\P}{\msf{P}}

\newcommand{\treeeval}{\textsc{TreeEval}}
\newcommand{\fullcirceval}{\textsc{FullCircuitEval}}
\newcommand{\circeval}{\textsc{CircuitEval}}

\newcommand{\bit}{\set{0, 1}}

\title{Improved Bounds on the Space Complexity of Circuit Evaluation}
\author{Yakov Shalunov\\University of Chicago\\\href{mailto:yasha@uchicago.edu}{yasha@uchicago.edu}}

\begin{document}
\maketitle

\begin{abstract}
    Williams (STOC 2025) recently proved that time-$t$ multitape Turing machines can be simulated using $O(\sqrt{t \log t})$ space using the Cook-Mertz (STOC 2024) tree evaluation procedure. As Williams notes, applying this result to fast algorithms for the circuit value problem implies an $O(\sqrt{s} \cdot \polylog s)$ space algorithm for evaluating size $s$ circuits.
    
    In this work, we provide a direct reduction from circuit value to tree evaluation without passing through Turing machines, simultaneously improving the bound to $O(\sqrt{s \log s})$ space and providing a proof with fewer layers of abstraction.

    This result can be thought of as a ``sibling'' result to Williams' for circuit complexity instead of time; in particular, using the fact that time-$t$ Turing machines have size $O(t \log t)$ circuits, we can recover a slightly weakened version of Williams' result, simulating time-$t$ machines in space $O(\sqrt{t} \log t)$.
\end{abstract}
\section{Introduction}
Recently, Williams proved an extremely counterintuitive result: 

\begin{theorem}[Williams~\cite{williams25}]\label{thm:time-in-space}
    All time-$t(n) \geq n$ multitape Turing machines can be simulated using $O(\sqrt{t \log t})$ bits of space.
\end{theorem}

This holds no matter how they use the $\Omega(t)$ cells of space they can touch and improves dramatically on the 50-year-old best-known simulation using space $O(t / \log t)$ due to Hopcroft, Paul, and Valiant~\cite{hpv77}.

This simulation can be composed with Turing machine algorithms for evaluating circuits to obtain space-efficient algorithms for circuit evaluation. In particular, the best known algorithm for circuit simulation on Turing machines is due to Pippenger~\cite{pippenger77} (Theorem~\ref{thm:fast-circuit}) and yields a space $O(\sqrt{s} \cdot \log^{3/2} s)$ algorithm for evaluating arbitrary size-$s$ circuits.

Motivated by the fact that circuits themselves are an appealing model of fine-grained complexity, we directly prove an improved bound on the space complexity of circuit evaluation:

\begin{theorem}[Main theorem]\label{thm:main}
    Given a size $s$ circuit $C$ on $n \leq s$ inputs and an input $x \in \bit^n$, the output $C(x)$ can be evaluated in space $O(\sqrt{s \log s})$.
\end{theorem}

The uniform formulation immediately implies a nonuniform version:
\begin{cor}
    If $f: \bit^n \to \bit$ can be computed by a size $s$ circuit then $f$ can be computed by a size-$2^{O(\sqrt{s \log s})}$ branching program.
\end{cor}

Additionally, Theorem~\ref{thm:main} together with standard results for simulating Turing machines with circuits (see Theorem~\ref{thm:small-circuit-for-time}) gives $\TIME[t] \subseteq \SPACE[\sqrt{t} \log t]$, re-proving Williams' result up to log factors.

\subsection{Technique outline}
Similarly to Williams', our result is based on a reduction to the tree evaluation problem. Roughly speaking, the tree evaluation problem (definition~\ref{def:treeeval}) is the task of evaluating a height $h$, $d$-ary tree where each leaf is a $b$-bit value and each internal node is function $\bit^{db} \to \bit^b$, given as an explicit table of $2^{db}$ $b$-bit entries.

The natural depth-first approach requires space $O(h d b)$ to evaluate the tree, storing $d b$ bits at each of $h$ recursive levels. Though this depth-first approach intuitively seems ``inherent'' to the problem (indeed, the problem was originally posed to attempt to separate $\L$ from $\P$), Cook and Mertz came up with an $O(h \log db + db)$-space algorithm for the problem, improving the space complexity quadratically when $h \approx d b$~\cite{cookmertz24}.\footnote{It is worth noting that several years prior to this $O(\log n \cdot \log \log n)$ result, Cook and Mertz first showed that the problem can be solved in space $O(\log^2 n/\log \log n)$~\cite{cookmertz20,cookmertz21}. (Here, $n = \tilde \Theta (d^h \cdot 2^{d b})$ so $\log n = \Theta (h \log d + db)$.)}

Williams' result reduces Turing machine computation to a tree evaluation instance of arity $d = O(1)$ with height $h = \Theta(t/b)$ and value-size $b$ for some parameter $b$---evaluating this instance naively earns nothing, since $\Theta\q(\frac{t}{b}) \cdot b = \Omega(t)$, but balancing $t/b$ and $b$ and then applying the Cook-Mertz tree evaluation procedure yields quadratic savings.

We provide a direct reduction from circuit evaluation to tree evaluation which bypasses reasoning about the motion of Turing machine heads. For intuition, we first sketch a warm-up result:

\begin{prop}
    Size $s$ circuits can be evaluated in space $\tilde O (s^{2/3})$.
\end{prop}
\begin{proof}[Proof sketch.]
    Similar to the approach used for Turing machines, we partition our circuit into blocks of size $b$. This, in particular, ensures that the \emph{number} of blocks (and thus depth) is $\frac{s}{b}$. Our ``blocks'' will be intervals of consecutive vertices under a topological ordering, thus ensuring that the quotient graph\footnote{Here, when we say ``quotient'' we mean a graph with a vertex for each part of the partition and a single edge between any two parts whose components have any edges in the original graph. A more formal definition can be found in the preliminaries in Section~\ref{sec:prelims}.} is a DAG.

    We can view this quotient DAG as a ``circuit'' over values from $\bit^b$ instead of over individual bits: the value at each vertex in the quotient graph is the list of values of all $b$ gates in the corresponding block of the original circuit and each wire carries all $b$ of those bits.

    Since the quotient DAG connects blocks together when any pair of underlying gates have a connection, all gates in the block have sufficient information to be evaluated and our ``quotient circuit'' is able to simulate the original circuit.

    Now we observe that in much the same way that one can expand a circuit into a formula, we can expand this quotient circuit into a tree evaluation instance (i.e., duplicating each vertex once for every path to reach it from the root). This increases the size exponentially (which is fine, because we'll never write down the whole tree, instead computing it on-demand) but, crucially, does not affect the depth. 
    
    Since the depth of the original quotient circuit was bounded by the number of blocks in it, which in turn was $\frac{s}{b}$, the height of our tree evaluation instance is $h = \frac{s}{b}$, and since each block has $b$ gates, the value-size parameter is exactly $b$.

    In order to get $\tilde O(\sqrt{s})$, we ultimately need an arity of $d = O(1)$; unfortunately the naive construction given here only allows a trivial bound on the in-degree: given a block $B$, each wire coming into $B$ (i.e., edges $(u, v)$ with $v \in B$) in the original circuit connects to one vertex and thus adds at most one block (the block $B(u)$ containing $u$) to the list of blocks with wires to $B$. Since there are $b$ gates and each has two inputs (we work with binary circuits), there are at most $2b$ total incoming wires, and thus at most $2b$ connected blocks.

    This gives us $d \approx b$ and so the Cook-Mertz procedure gives us space $O(\frac{s}{b} \log b^2 + b^2)$. Setting $b \approx \sqrt[3]{s}$ gives $\tilde O(s^{2/3})$.
\end{proof}

In order to get $\tilde O (\sqrt {s})$, we will need the key technical contribution of this work: a trick for reducing the in-degree of the quotient graph, expressed (in slightly simplified form\footnote{The full version is stated as Lemma~\ref{lemma:dag-partition-full}; it provides additional complexity guarantees and allows for trading off between the final in-degree and depth. (In particular, a final in-degree $d + 1$ allows a depth of $s/b$.)}) in the following graph-theoretic lemma:
\begin{lemma}[Low-degree DAG partition (simplified)]\label{lemma:dag-partition}
    For any directed acyclic graph $G$ of size $s$ and maximum in-degree $d > 1$ and any integer choice of $b \in [d, s]$, there is a subdivision\footnote{A subdivision of a graph subdivides some edges by inserting vertices into them (i.e., replacing $e = (u, v)$ with a vertex $v_e$ and edges $(u, v_e), (v_e, v)$). In the context of circuits specifically, these will be additional identity gates spliced into some wires, which trivially does not affect behavior.} $G'$ of $G$ and a partition $P$ of $G'$ such that:
    \begin{itemize}
        \item Every part in $P$ has at most $b$ vertices.
        \item Every vertex in the quotient graph $G' / P$ has in-degree at most $2$.
        \item All directed paths in $G' / P$ have length less than $d \cdot \frac{s}{b}$.
    \end{itemize}
\end{lemma}
Since we consider binary circuits, our application will have $d = 2$, and $b = \tilde \Theta (\sqrt{s})$, yielding a depth of $2 \cdot \frac{s}{b} - 1 = \tilde O(\sqrt{s})$ and giving us the desired result.

Intuitively, we are ``buying'' a lower depth by ``paying'' with increased block size---were we doing a naive depth-first evaluation, this would get us nothing, but using Cook-Mertz tree evaluation procedure allows us to have increased block size ``for free'' (up to the point where it exceeds depth). 

We remark that when we reduce the depth of the graph in the proof of the lemma, we do so by ``brute-force,'' of a sort: we first create a quotient graph on the original graph $G$ which has a total \emph{size} (and thus depth) of $d \cdot \frac{s}{b}$ and then use subdivision to reduce the in-degree without increasing the depth. This allows us to reduce the depth of an \emph{arbitrary} graph without care for its structure since the depth of the quotient graph cannot possibly exceed its size.

\subsubsection*{Outline}
In section \ref{sec:prelims}, we give some preliminaries on the circuit and tree evaluation problems. In section \ref{sec:partition}, we prove the partition lemma (Lemma~\ref{lemma:dag-partition}). Finally, in section \ref{sec:reduction}, we apply the partition lemma to reduce circuit evaluation to an appropriate tree evaluation instance.

% The latter result follows immediately from two facts about oblivious Turing machines. (An ``oblivious'' Turing machine is one whose head position at any given time step $i$ depends only on the input length $n$ and not the specific input $x$.) The first result is as seen in~\cite{williams25}.
% \begin{theorem}[\cite{hennie66, pippengerfischer79, lance05}]
%     For every deterministic, time $t(n)$ multitape Turing machine $M$, there exists an equivalent oblivious Turing machine running in time $O(t \log t)$. Further, the head positions at time $i$ can be computed in time $\polylog t$.
% \end{theorem}
% \begin{theorem}[\cite{heribert10}]
%     If $L$ is decided by an oblivious Turing machine running in time $t(n)$ whose head positions can be efficiently computed then $L$ can be computed by uniform circuits of size $O(t)$.\footnote{The original result does not argue any uniformity, but the construction is a modified tableau which uses the knowledge of the head positions to add only constant number of gates (corresponding to the cell under the head) per time step; correspondingly, if the head positions are computable in space $S$, the circuit is constructible in space $O(S + \log t)$.}
% \end{theorem}

\section{Preliminaries}\label{sec:prelims}
\paragraph{Model relationships}
First, the following pair of results together illustrate the close connection between circuits and Turing machine time:
\begin{theorem}[Pippenger~\cite{pippenger77}]\label{thm:fast-circuit}
    Given a size $s$ circuit $C$ and an input $x$, the output $C(x)$ can be evaluated in multitape Turing machine time $O(s \log^2 s)$.
\end{theorem}
\begin{remark}
    Pippenger's proof of this result does not appear to be available online; correspondingly, we have included an exposition of the algorithm in appendix~\ref{app:fast-circuit-eval}. See also Williams' discussion of the result~\cite{williams25}.
\end{remark}
\begin{theorem}[Pippenger, Fischer \cite{pippengerfischer79}, Kouck\'y~\cite{kouckyreport}]\label{thm:small-circuit-for-time}
    For every $t(n) \geq n$ and $L \in \TIME[t]$, the $\log$-space-uniform circuit complexity of $L$ is $O(t \log t)$.
\end{theorem}
\begin{remark}
    While this result is originally due to Pippenger and Fischer (proven by passing through an oblivious Turing machine simulation), Michal Kouck\'y developed a simpler direct proof~\cite{kouckyreport}; this result appears not to be available digitally either and has been reproduced in appendix~\ref{app:small-tm-circuits}.
\end{remark}

Additionally, we follow Williams and remark that unlike for time complexity (where best-known simulations incur polylogarithmic or even quadratic overhead) the case of space complexity is much cleaner: all standard models (e.g., single tape, multitape, and random access Turing machines as well as register-based random access machines) can be mutually simulated with only constant-factor space overhead~\cite{emdeboas1990,slot1998,williams25}. Correspondingly, we do not worry about the exact computational model underlying our algorithms.

\paragraph{Circuits} Formulations of the circuit evaluation/circuit value problem differ. However, all that we are aware of reduce trivially to the $\fullcirceval$ formulation below. The following lemma justifies working with the simplified $\circeval$.
\begin{define}[Circuit Evaluation]
    $\fullcirceval$ instances take the form $(x, C)$ where $x$ is an $m$-bit string and $C$ is an $s$-gate circuit in the full binary basis on $m$ inputs, provided as a list of triples of the form $(\ell_i, r_i, \phi_i)$ where $\phi_i: \bit^2 \to \bit$ is a binary function and each of $\ell_i$ and $r_i$ is either a gate index $j \neq i$ or a variable index $k$ for some $k \leq m$.

    The objective is to output the value $v_i$ of each gate. For simplicity, assume $s \geq m$.

    A $\circeval$ instance $C$ is an $s$-gate circuit in the full binary basis provided as a list of triples of the form $(\ell_i, r_i, \phi_i)$ where $\phi_i: \bit^2 \to \bit$ is a binary function and each of $\ell_i$ and $r_i$ are either a gate index $j < i$ (i.e., the input is in topologically sorted order) or a constant bit.

    The objective is to output the value of the last gate.
\end{define}

\begin{lemma}[$\circeval$ is good enough]
    If there is a space $S$ algorithm for $\circeval$ then there is a space $S + \log^2 s$ algorithm for $\fullcirceval$.
\end{lemma}
\begin{proof}
    First, observe that given an instance of $\fullcirceval$, we can replace every reference to variable $k$ with the value $x_k$ in space $\log s$ by simply scanning over the gates of $C$ and for each one, if $\ell_i$ is a gate index, outputting it as is, and if it is a variable index $k$ then storing the current gate in $\log s$ bits and the variable index in $\log m \leq \log s$ bits and scanning back to find $x_k$, outputting that value instead of $k$. Similarly for $r_i$.

    Next, observe that it is possible to topologically sort a graph in $O(\log^2 s)$ space~\cite{cook85}.

    Finally, suppose $M$ computes $\circeval$ in space $S$. We use space-efficient composition and the above procedures to run $M$ on gates $\set{0, \dots, i}$ for each $i$, sorting with respect to the sink $i$. Between computations we only need to store which gate we are on using $\log s$ bits and everything else can be reused.
\end{proof}

\paragraph{Graphs} When we refer to a quotient graph with respect to a partition, we refer to a graph which collapses together all vertices inside a part of the partition and then deduplicates all the edges. That is, if $G = (V, E)$ is a graph and $P: V \to [k]$ is a partition into $k$ parts, we say that for $i, j \in [k]$, parts $i, j$ are connected in $G/P$ if there exist $u, v \in V$ such that $P(u) = i$, $P(v) = j$, and $(u, v) \in E$. There is at most one edge between any pair of parts and we will discard self-loops. We will also use the term ``block'' to refer to parts of the partition.

Given a graph $G = (V, E)$, an ``edge subdivision at $e$'' is the operation of taking an edge $e = (u, v) \in E$ and removing it, adding in its place a vertex $v_e$ and edges $(u, v_e)$ and $(v_e, v)$. The subdivided graph then has $V' = V \cup \set{v_e}$ and $E' = E \cup \set{(u, v_e), (v_e, v)} \without{e}$. A subdivision $G'$ of $G$ is any graph created by 0 or more edge subdivisions.

For the purposes of the complexity of operations, we make some comments about representations of objects: partitions are specified as an explicit function mapping vertices to the labels of their part and the set of labels of vertices in a subdivided graph is a superset of the labels in the original graph. This allows us to maintain the gate information of the circuit and to identify all subdivision vertices to assign them identity gates.

\paragraph{Tree evaluation} Formally, the tree evaluation problem is defined as follows:

\begin{define}[Tree Evaluation]\label{def:treeeval}
    $\treeeval$ instances are full $d$-ary trees of height $h$ where each leaf $\ell \in \set{0, \dots, d-1}^h$ is labeled with a $b$-bit value $v_\ell$ and each internal node $u \in \set{0, \dots, d-1}^{<h}$ is labeled with an explicit function (provided as a table of values) $f_u: \bit^{d b} \to \bit^b$.

    We recursively define the value $v_u$ at each internal node $u$ in the natural way to be $$v_u := f_u (v_{u0}, \dots, v_{u(d-1)})$$. The objective is to evaluate $v_\mrm{root} \in \bit^b$, the value of the root.
\end{define}
(For the history and broader significance of the tree evaluation problem, see Cook and Mertz work~\cite{cookmertz24}. We use the formulation of the problem exposited by Goldreich~\cite{goldreich24}.)
\begin{theorem}[Cook-Mertz~\cite{cookmertz24, goldreich24}]\label{thm:treeeval}
    $\treeeval$ instances of height $h$, arity $d$, and value-size $b$ can be evaluated in space $O(h \log db + db)$.
\end{theorem}

\section{Graph partitioning}\label{sec:partition}
First, let us state the lemma in its general form:
\begin{lemma}[Low-degree DAG partition]\label{lemma:dag-partition-full}
    For any directed acyclic graph $G$ of size $s$ and maximum in-degree $d > 1$, for any integer choice of parameters $d' \in [2, d + 1]$ and $b \in [d/d', s]$, there is a subdivision $G'$ of $G$ and a partition $P$ of $G'$ such that:
    \begin{itemize}
        \item Every part in $P$ has size at most $b$.
        \item Every vertex in the quotient graph $G' / P$ has in-degree at most $d'$.
        \item $G' / P$ is a layered DAG with layer count at most $\frac{d}{d' - 1} \cdot \frac{s}{b}$.
        \item Given $G$ in topologically sorted order, $d'$, and $b$, the subdivision $G'$ and partition $P$ can be computed in $O(\log s)$ space.
    \end{itemize}
\end{lemma}

\begin{remark}
    By allowing an additional ``vertex subdivision''\footnote{That is, replacing a $v$ with two vertices $v_\mrm{in}$ and $v_\mrm{out}$ which are connected by an edge $(v_\mrm{in}, v_\mrm{out})$ and replacing all edges $(u, v)$ with edges $(u, v_\mrm{in})$ and edges $(v, w)$ with $(v_\mrm{out}, w)$. In the context of circuits, $v_\mrm{in}$ is the original gate and $v_\mrm{out}$ is a forwarding identity gate.} operation, one can obtain a slightly stronger version of the lemma, obtaining optimal depth $s/b$ when $d = d' = 2$. However, the factor-of-$2$ savings on depth do not change the bottom line result (Theorem~\ref{thm:main}), the construction is somewhat more convoluted, and it is not immediately clear how it generalizes to arbitrary $d$ and $d'$.

    We also believe it should be possible to improve the $\frac{d}{d' - 1}$ factor to either $\frac{d - 1}{d' - 1}$ or $\frac{d}{d'}$ without the introduction of a new operation, but this seems to significantly increase the complexity of the proof.
\end{remark}

As in the statement of the lemma, let $G$ be a size-$s$ DAG of max in-degree $d$ and let $d' \in [2, d + 1]$, $b \in [d/d', s]$ be parameters.

\newcommand{\defsym}[1]{#1}
\paragraph{Symbols and indices} Throughout the proof: \defsym{$s$} refers to the size of the input graph $G$; \defsym{$d$} is the max in-degree of $G$; \defsym{$d'$} is the target in-degree of the quotient graph; \defsym{$b$} is the maximum size of blocks and the value-size in the tree evaluation instance; \defsym{$b_0$} refers to the size of the initial blocks; \defsym{$i$} and \defsym{$j$} refer to indices of blocks or directly comparable values; \defsym{$\ell$} is used to index incoming edges in the quotient graph; finally \defsym{$t$} refers to the index of the final initial block and is the depth of the graph.

\subsection{Construction}

\paragraph{Initial partition} Though ultimately we produce a subdivision and a partition of that subdivision, we start with a partition of the original graph. We will use this initial partition to describe the subdivision and ultimate partition. In particular, the mappings of the vertices in $G$ will in the final partition will be the same as the initial partition: we will simultaneously add subdivision vertices and blocks to contain them.

While blocks in the ultimate partition are allowed size up to $b$, the initial blocks have to be somewhat smaller: specifically, they will be size $b_0 = \floor{\frac{d' - 1}{d} \cdot b}$. Let $V = (v_0, \dots, v_{s-1})$ be a topological ordering of $G$ and let $t = \ceil{s /b_0} - 1$. Then for $i < t$, we define $$
    B_i := (v_{i b_0}, \dots, v_{(i + 1)b_0 - 1})
$$
where for the final block $B_t$ we truncate to $v_{t b_0}, \dots, v_{s - 1}$. In the notation of a mapping of vertices to parts, this is the mapping $V = \set{0, 1, \dots, s-1} \to \set{0, 1, \dots, t}$ given by $k \mapsto \floor{k/b_0}$. Note that the resulting quotient graph is a DAG and the ordering $B_i$ is a topological sort since if $i < j$, $v_k \in B_i$ and $v_{k'} \in B_j$ then $k < k'$. Thus, there is no edge $(k', k)$ and so there can be no edge $(j, i)$.\footnote{Note that \emph{any} partition $P$ of $G$ such that $G / P$ is a DAG must have the property that each block in $P$ represents an interval in \emph{some} topological sort.}

Observe that we have chosen $b_0$ such that each block has at most $b_0 \cdot d = b \cdot (d' - 1)$ incoming edges in the original graph. As some intuition: in order to manage the in-degree of the quotient graph, we will subdivide incoming edges and group the newly created vertices into $d' - 1$ blocks which funnel them into the block (with the final $d'$th block being the immediately preceding block).

As in the sketch, we have successfully created a quotient DAG where the maximum depth is $t + 1 = \frac{s}{b_0} = \frac{d}{d' - 1} \cdot \frac{s}{b}$ and all parts have size at most $b$. Now we must address the key issue: the in-degree. The source vertices of those $b \cdot (d' - 1)$ edges have no reason to be confined to $d'$ blocks.\footnote{In fact, it is possible to construct graphs where this procedure will yield max in-degree of $\min(d' \cdot b, t - 1)$ (i.e., every incoming edge leads to a different block, as long as enough blocks exist for that). As an explicit example, consider $d = 2, d' = 3$, and $s = b^2$, with edges $(k, k +1)$ for all $0 \leq k < s -1$ and edges $(k,tb + k/b)$ for $k < t$ where $b | k$. That is, a graph where a ``spine'' of edges enforces a specific topological ordering and then the first vertex of each block $B_i$ connects to vertex $i$ in block $B_t$. Then $B_t$ has in-degree $t - 1 \gg d'$.}

\paragraph{Cables} In order to resolve this, we will introduce ``cables'' to ``gradually gather together'' the incoming edges into a small number of blocks. For every edge $e = (u,v)$ with $u \in B_i$ and $v \in B_j$ for $i < j$, we will subdivide $e$ $j - i - 1$ times, creating vertices $v_{e,k}$ for $1 \leq k < j - i$. This gives us our graph $G'$. 

Now we need to partition the newly-created vertices. For simplicity, let us first consider the case where $d' = 2$. To get the desired result: for every edge $e$ starting outside $B_j$ and terminating in $B_j$, we will add the vertex $v_{e,k}$ (if it exists) to block called $B_{j,k}$ for $k < j$. Since, in this case, there are at most $d b_0 = b \cdot (d' - 1) = b$ edges $e$ terminating in $B_j$, the size of each block $B_{j, k}$ is at most $b$. 
\begin{figure}
    \centering
    \begin{tikzpicture}[every node/.style={circle, draw, minimum size=3em, inner sep=0, scale=0.75},]
        \path 
            (0,0) node (A) {$B_0$}
            (0.25,1.35) node (B) {$B_1$}
            (1.4,3.5) node (C) {$B_{j - 2}$}
            (2.6,4.7) node (D) {$B_{j - 1}$}
            (4,5.7) node (E) {$B_j$}
        ;
        \draw[->] (A) -- (B);
        \draw[->, dashed] (B) -- (C);
        \draw[->] (C) -- (D);
        \draw[->] (A) to[out=0,in=265] (E);
        \draw[->] (B) to[out=10,in=250] (E);
        \draw[->] (C) to[out=10,in=235] (E);
        \draw[->] (D) -- (E);
    \end{tikzpicture}%
    \hspace{5em}%
    \begin{tikzpicture}[every node/.style={circle, draw, minimum size=3em, inner sep=0, scale=0.75},]
        \path 
            (0,0) node (A) { $B_0$}
            (0.25,1.35) node (B) { $B_1$}
            (1.4,3.5) node (C) { $B_{j - 2}$}
            (2.6,4.7) node (D) { $B_{j - 1}$}
            (4,5.7) node (E) { $B_j$}
            (2.8,1.2) node (C1) { $B_{j,j-1}$}
            (3.3,2.4) node (C2) { $B_{j,j-2}$}
            (3.9,4.4) node (C3) { $B_{j, 1}$}
        ;
        \draw[->] (A) -- (B);
        \draw[->, dashed] (B) -- (C);
        \draw[->] (C) -- (D);
        \draw[->] (D) -- (E);
        \draw[->] (A) to[out=5,in=230] (C1);
        \draw[->] (B) to[out=5,in=210] (C2);
        \draw[->] (C) to[out=5,in=220] (C3);
        \draw[->] (C1) -- (C2);
        \draw[->, dashed] (C2) -- (C3);
        \draw[->] (C3) -- (E);
    \end{tikzpicture}% 
    
    \caption{Addition of ``cable'' blocks to reduce in-degree of block $B_j$ in the $d' = 2$ case. In the full construction, the other $B_i$ blocks have their own cables, which are omitted here for clarity.}
    \label{fig:cable-blocks}
\end{figure}

As suggested by the notation $B_{j,k}$, the blocks $B_{j,1}$ through $B_{j, j - 1}$ form a sort of ``tail'' or ``cable'' connected to $B_j$; they run along the preceding blocks of the graph, connecting to each one at the appropriate point in the cable (specifically, block $B_i$ is connected to the cable at block $B_{j, j - i - 1}$). The cable carries all the edges terminating in block $B_j$ from earlier blocks (see Figure~\ref{fig:cable-blocks}).

Now we consider the case where $d' > 2$. The ``bandwidth'' of the cable is $b$, and that becomes insufficient when there are up to $(d' -1)b$ incoming edges. We solve this problem in a very natural way: add more cables. Formally: in the case where $d' > 2$, we will instead have blocks $B_{j,k}^\ell$ where $k < j$ indexes the ``layer'' as before and $1 \leq \ell < d'$ indexes the cable. We can enumerate the edges $\set{e_1, \dots, e_{b \cdot (d' - 1)}}$ coming into $B_j$. Then $v_{e_r, k}$ will be placed into block $B_{j,k}^{\floor{r/b}}$.

For ease of computation, we will define this enumeration of the edges to be indexed by first the index of the terminal vertex and then by which of $\leq d$ incoming edges it corresponds to (sorted by whatever order they appear in in the input representation). We will allow indices to be skipped: for example, $e_d$ will be reserved for the $d$th incoming edge of vertex $v_{j b}$, even if the in-degree of $v_{j b}$ is strictly less than $d$ and that edge does not exist. This completes the description of the partition $P$ of $G'$.

\subsection{Analysis}\label{subsec:analysis}
Recall that we want the following properties:
\begin{itemize}
    \item Every part in $P$ has size at most $b$.
    \item Every vertex in the quotient graph $G' / P$ has in-degree at most $d'$.
    \item $G' / P$ is a strictly layered DAG with depth (i.e., layer count) at most $\frac{d}{d' - 1} \cdot \frac{s}{b}$.
    \item Given $G$ in topologically sorted order, $d'$, and $b$, the subdivision $G'$ and partition $P$ can be computed in $O(\log s)$ space.
\end{itemize}

\begin{description}
    \item[Size] The first condition is trivially satisfied for the initial blocks which have size $b_0 = \frac{d' - 1}{d} b \leq b$. It is satisfied for the cable blocks since each cable block gets at most $b$ gates assigned to it because the map $[b \cdot (d' - 1)] \to [d' - 1]$ given by $r \mapsto \floor{r/b}$ is $b$-to-1.

    \item[In-degree] The second condition is satisfied for the cable blocks since each $B_{j,k}^\ell$ has incoming edges only from $B_{j,k+1}^\ell$ and $B_{j - k - 1}$ and is satisfied by construction for the initial blocks since $B_j$ has incoming edges only from $B_{j,1}^\ell$ for $\ell \in [d' - 1]$ and from $B_{j - 1}$.

    \item[Layering] The third condition can be seen similarly to the second: there are $t + 1$ layers $L_0, \dots, L_t$ where $B_j$ belongs to $L_j$ and $B_{j,k}^\ell$ belongs to $L_{j - k}$. Since each block $B_j \in L_j$ has incoming edges from only $B_{j,1}^\ell \in L_{j - 1}$ and $B_{j - 1} \in L_{j - 1}$ and each block $B_{j,k}^\ell \in L_{j - k}$ connects only to $B_{j - k - 1} \in L_{j -k -1}$ and $B_{j,k + 1}^\ell \in L_{j - k - 1}$. Thus, blocks in $L_i$ connect only to blocks in $L_{i + 1}$, and so the partition $\set{L_0, \dots, L_t}$ represents a ``strict layering.'' In particular, this trivially implies that the maximum directed path length is at most $t = \frac{s}{b_0} - 1 = \frac{d}{d' - 1} \cdot \frac{s}{b} -1$.

    \item[Complexity] First, note that the parameter $b_0$ can be computed from $b$, $d$, and $d'$. Since $d$ is not given, it needs to be computed from the graph $G$ but this can be done by iterating over all vertices and keeping a running max of their in-degrees (which in turn can be computed by iterating over all other vertices and counting how many have an edge to the given vertex).
    
        The computation of the subdivision is arithmetic: for each edge $e = (v_m, v_n)$ we compute $i = \floor{m / b_0}$ and $j = \floor{n / b_0}$ and then subdivide that edge $j - i - 1$ times. Since this is all arithmetic on indices, it is logspace computable. 
    
        The computation of the partition is similarly efficient: $v_m$ is assigned to $B_{\floor{m / b_0}}$ and $v_{e, k}$ for $e = (v_m, v_n)$ is assigned to $B_{j, k}^\ell$ where $\ell = \floor{r/b}$ and $r = d \cdot (n \mod b_0) + p$ where $p$ is the index of $e$ among the incoming edges of $v_n$ (which is logspace computable since we define the ordering on the edges for a given sink vertex to be whatever order they appear in in the input representation). 
        
        When outputting the subdivision, we can trivially output all vertices of the original graph first followed by the subdivision vertices, satisfying the condition that labels of vertices be preserved.
\end{description}
\qed

\section{Circuit evaluation reduction}\label{sec:reduction}
Our reduction to tree evaluation can be stated as follows:
\begin{theorem}
    Given a topologically sorted circuit of size $s$ and any $b \in [2, s]$, we can produce an equivalent $\treeeval$ instance\footnote{Technically, the $\treeeval$ instance outputs $b$ bits while the circuit outputs 1; we actually produce a $\treeeval$ instance where the output of the circuit is bit $s \mod b$ of the output.} with arity $d = 2$, height $h = O(\frac{s}{b})$ and word-size $b$ as chosen. Further, this reduction is locally computable in space $O(b + \log s)$. 
    
    That is, there is a space $O(b + \log s)$ procedure which, on input $(C, b, u, x, y)$ (where $C$ is a circuit, $b \in [2, s]$, $u \in \bit^{\leq h}$ specifies a node in the tree, $x,y \in \bit^b$) computes $f_u (x, y)$ where $f_u$ is from the $\treeeval$ instance equivalent to $C$.

\end{theorem}
In particular, choosing $b = \sqrt{s \log s}$ and applying Theorem \ref{thm:treeeval} immediately gives Theorem \ref{thm:main}; further, a space $O(h + b)$ procedure for $\treeeval$ would immediately imply a space $O(\sqrt{s})$ procedure for $\circeval$.

At a high level, we apply the DAG partition lemma (vertices introduced by subdivision become identity gates) to $C$ and expand $C' / P$ into a tree, letting the function $f_B (x, y)$ computed at a node $B \in C' / P$ be simply the subcircuit $B$ evaluated with the external input wires taken from $x, y$. Note that $B$ may be smaller than size $b$ and may have fewer than $2b$ external input wires. We can embed each part into a value by saying that the values of gates fill the bits in the $\bit^b$ values in topologically sorted order.

Further, the part of computing $f_B (x, y)$ corresponding to actually evaluating the subcircuit $B$ can be done by naively evaluating each gate in topological order from previous ones and writing it down, yielding the $O(b)$ component of the space.

It turns out that there are no pitfalls, and the above description just works. Nonetheless, it is stated somewhat more rigorously below for completeness.
\begin{proof}
    First, note that we can efficiently identify the children of a given block (e.g., by iterating over all blocks and checking whether they are connected, which can in turn be done by iterating over all pairs of vertices in the two blocks being considered). We can identify the root block by finding which block the vertex $v_{s - 1}$ maps to. We will consider the ``left'' child of a block to be the child block whose label is lexicographically first and the ``right'' child to be the one whose label is lexicographically second. 

    Given as input a circuit $C$ in the full binary basis, the parameter $b$, and $(u, x, y)$, using space-efficient composition, we:
    \begin{enumerate}
        \item Apply the DAG partition lemma with $b$ as given and $d' = 2$. Initialize $B \gets \msf{root}$.
        \item For each bit $u_i$ of $u$, identify the left or right child of $B$ based on whether $u_i = 0,1$ and update $B$ to that child. 
        \item Once we have found the block $B$ corresponding to the tree node $u$, we must evaluate it. If it is a leaf block, it must have no inputs, so we compute it by naive dynamic programming.\footnote{The space complexity of the reduction could be improved to $O(\sqrt{b \log b} + \log s)$ by recursing, but this would not improve the bottom line space complexity of circuit evaluation since the space is already dominated by the evaluation of the tree evaluation instance.}
        
        Otherwise, since we defined values to fill the $b$-bit values in topologically sorted order, for each input $v \in B_c$ (for $c \in \set{x, y}$) to a gate $w \in B$, we iterate over $G'$ and count how many gates we find in $B_c$ before reaching $v$ (call it $\ell$); we then take bit $c_\ell$ as the corresponding input to $w$. This allows us to locally compute the circuit $B$ with external inputs substituted in.
        \item Having computed (or rather, expressed a procedure by which we can space-efficiently compute bits of) the circuit $B$ and which bit of $x, y$ each input to $B$ corresponds to, we compute the values of $B$ via naive dynamic programming.
    \end{enumerate}

    In order to extract the value of the circuit from the reduction, we then ultimately take bit $s \mod b_0$ of $v_\msf{root}$, since that is the location of the last gate in the final block.

    The correctness of the reduction is straightforward: the behavior of the circuit $C'$ is the same as $C$ since we just inserted some identity gates into wires. Further, the value of a block $B \in C' / P$ can be computed using the values of all the gates some gate in $B$ takes input from, which by construction are just the 2 children of $B$.

    For complexity, observe that on-demand computing bits of $G'$ and $P$ as needed takes space $O(\log n)$; step 2 requires holding a label $B$ and index $i$ into $u$, which also takes space $O(\log n)$; iterating over $G'$ in step 3 requires a single log-sized counter and counting gates in $B_c$ similarly requires logspace since we just compute for each gate which block it belongs to (which we are given explicitly) and maintain a counter of how many have been in $B_c$; step 4 is done via naive dynamic programming, which requires space $O(b)$ to evaluate a size-$b$ circuit.
\end{proof}

\section{Acknowledgments}
I would like to thank William Hoza for suggesting this research direction, helping review this article, and general guidance. I am also grateful to Alexander Razborov for feedback on presentation.

\bibliographystyle{alpha}
\bibliography{bib}

\appendix

\section{Fast circuit evaluation}\label{app:fast-circuit-eval}
Because it is not digitally available, we provide an exposition of Pippenger's algorithm below. We make no claims of originality. Note that this exposition is not intended to be rigorous so much as explanatory, so we have omitted a proof of correctness or runtime analysis.
\begin{theorem*}[Pippenger~\cite{pippenger77}]
    Given a size $s$ circuit $C$ and an input $x$, the output $C(x)$ can be evaluated in multitape Turing machine time $O(s \log^2 s)$.
\end{theorem*}

We will first describe the high-level recursive algorithm; in order to implement it on a multitape Turing machine, one simply needs to create a tape for each ``local variable'' and treat this tape as a stack, pushing the new value of each local variable to its stack when recursing. By, for example, using a marker to separate layers of the stack, any ``well-behaved'' recursive algorithm (i.e., one which does not try to read back up into the caller's stack frames) can be expressed using a constant number of tapes, since the ``current level'' of each stack can function as an arbitrary (one-sided) Turing machine tape.

We will work with lists of ``records,'' which will be various tuples of indices (i.e., all entries in the list have the same size which is $O(\log s) = O(\log(\text{input length}))$). We will use the following high-level operations:
\begin{description}
    \item[Classify] Given a list of $n$ records and a linear-time predicate, ``classify'' the records into two lists based on whether they satisfy the predicate. On a multitape Turing machine, we can perform this operation in time $O(n \log s)$ trivially if the source list and two destination lists are all on separate tapes.
    \item[Merge] Given a linear-time comparison operation and two lists of lengths $n$ and $m$ records sorted according to the comparator, ``merge'' the two lists into one sorted list of length $n + m$. On a multitape Turing machine, we can perform this operation in time $O((n + m) \log s)$ if the source lists and destination list are all on different tapes.
    \item[Sort] Given a list of $n$ records and a linear-time comparator, sort the list of $n$ records according to the comparator. On a multitape Turing machine, this can be done recursively with a constant number of auxiliary tapes in time $O(n \log n \log s)$.
\end{description}

Much like in the $\fullcirceval$ problem, we will assume that the input is a pair $(x, C)$ of input and circuit, with the circuit encoded as a list of $s$ records. However, we will assume that $C$ is provided in topologically sorted order. We will adopt the convention that $0 < 1 < x_1 < \dots < x_n < v_1 < \dots < v_s$. Note that unlike in the space-bounded computation case, we cannot simply assume that the caller only wants the last bit, since rerunning the algorithm for each prefix of the circuit would make the running time quadratic. Correspondingly, the algorithm outputs the value of all gates.

The first step is to eliminate the input $x$ and bake it into the circuit as constants: first, we copy $x$ and $C$ to separate tapes. In the process of copying $C$, we will replace each record $(\ell_i, r_i, \phi_i)$ with the record $(\ell_i, r_i, \phi_i, i)$ (which we can do by keeping the counter $i$ on an auxiliary tape). 

Then we sort $C$ by the field $\ell$. Then scan $C$ and $x$ in parallel, advancing the $x$ head whenever $\ell$ increases, substituting the appropriate constants for the left inputs of each gate which takes a variable input. We can then repeat sorting by field $r$ to eliminate the rest of the inputs and, finally, sort $C$ by the index field to restore the original order.

At this point, we can copy $C$ back to the original input tape and discard the contents of all other tapes.

Now, we will convert the circuit to (a variant of) Pippenger's ``alternate representation''---instead of a list of gate records which each store their inputs, we will have a list of gate records, which are now just the index and gate function, together with a second list of ``wire''/``value'' records of the form $(u, v, d)$ where $u$ is either a gate index (wire) or constant (value), $v$ is a gate index, and $d \in \set{L, R, O}$ indicates whether this wire is the left input to $v$, the right input, or an ``output record.'' In the case of an output record, $u$ will be either $*$ (wire) or a constant (value), representing either a place holder for the value of the gate or the computed value of the gate.

We can do this by simply scanning over the circuit and for each gate $v = (\ell, r, \phi, i)$ outputting wire records $(\ell, i, L)$, $(r, i, R)$, and $(*, i, O)$ and gate record $(\phi, i)$.

Given sets of gates $I, J$, let $W_{I \to J}$ represent the set of wires starting in $I$ and terminating in $J$. Let $V_{I \to J}$ be the set of value records corresponding to those wires. In both cases let $I \to$ and $\to J$ refer to wires/values starting in $I$ and terminating in $J$ respectively. Let $G_I$ denote the gate records of $I$.

In the following procedure, each expression like $G_K$ or $V_{\to I}$ should be interpreted as the name of a local variable. Performing the Turing machine construction by replacing each variable with a tape interpreted as a stack of tape yields a multitape Turing machine with around 20 tapes (including a couple work tapes in addition to the stacks).

We now describe a recursive procedure which, given an interval $K$, $G(K)$, $V_{\to K}$, and $W_{K \to}$ computes $V_{K \to}$. If $\abs{K} = 1$, we simply evaluate the single gate in $K$ using the two value records in $V_{\to K}$ and then substitutes that value into each record in $W_{K \to}$. Otherwise:
\newcommand{\push}{\Leftarrow}
\begin{itemize}
    \item Split $K$ into $I$ and $J$, each half the size of $K$.
    \item \begin{itemize}
        \item Classify $G_K$ into $G_I$ and $G_J$.
        \item Classify $W_{K \to}$ into $W_{I \to}$ and $W_{J \to }$.
        \item Classify $V_{\to K}$ into $V_{\to I}$ and $V_{\to J} \setminus V_{I \to J}$.
    \end{itemize}
    \item Letting $K \gets I$, recurse to compute $V_{I \to}$ (which we move from the output stack $V_{K \to}$ to the stack $V_{I \to}$).
    \item Classify $V_{I \to}$ into $V_{I \to J}$ and $V_{I \to} \setminus V_{I \to J}$.
    \item Merge $V_{I \to J}$ with $V_{\to J} \setminus V_{I \to J}$ to produce $V_{\to J}$.
    \item Letting $K \gets J$, recurse to compute $V_{J \to}$ (which we again move to stack $V_{J \to}$).
    \item Merge $V_{I \to}$ and $V_{J \to}$ into $V_{K \to}$
\end{itemize}

In order to compute our final output, we ensure $V_{K \to}$ is sorted by output gate, and then scan over it and extract the bits $b$ from the records of the form $(b, v, O)$.

\section{Small circuits for Turing machines}\label{app:small-tm-circuits}
\begin{theorem*}[Pippenger, Fischer \cite{pippengerfischer79}, Kouck\'y~\cite{kouckyreport}]
    For every $t(n) \geq n$ and $L \in \TIME[t]$, the $\log$-space-uniform circuit complexity of $L$ is $O(t \log t)$.
\end{theorem*}
The construction proving this is due to Kouck\'y~\cite{kouckyreport}. We make no claims of originality and present it for completeness because it is a beautiful argument which seems not to be well known.

We first consider a single-tape machine $M$ as a warm-up. We will recursively construct a circuit $C_t$ which given a $t$-cell configuration with the head somewhere in the middle third simulates $M$ on that tape for $t/3$ steps and output the final $t$-cell configuration. We will repeat the following 3 times, advancing the simulation by $t/9$ steps each time:
\begin{enumerate}
    \item Subdivide the tape into 9 equally-sized blocks.
    \item Identify the index $i \in \set{2, 3, \dots, 8}$ such that the head is in block $i$.
    \item Swap blocks $i - 1, i, i +1$ with blocks $1, 2, 3$
    \item Apply $C_{t/3}$ to blocks $1, 2, 3$ (simulating $M$ on that subsection of the tape for $t/9$ steps)
    \item Swap blocks $i - 1, i, i + 1$ with blocks $1, 2, 3$
\end{enumerate}
(For the recursive base case, we fall back to the standard tableau argument for $M$ when $t$ is bounded by an arbitrary constant.)

Note that we are guaranteed that the tape head is in block $i$ for $i \in \set{2, 3, \dots, 8}$ on all three layers since the head starts in blocks $4, 5, 6$ and only moves at most one block per iteration, so at the start of the second layer it is in $3, \dots, 7$ and in the third layer it is in $2, \dots, 8$ as desired. Thus, the circuit is correct by induction.

The gadgets required for steps 2, 3, and 5 are linear size, so the overall size is $s(t) = 3s(t/3) + O(t) = O(t \log t)$.

In order to simulate $M$ running in time $t(n)$ on an input of length $n$, we take $C_{3t(n)}$, hard coding the left $t(n)$ and right $2 t(n) - n$ cells to be 0 (i.e., we place the input and thus the initial location of the head in the middle $t(n)$ cells). 

In order to simulate a multitape Turing machine, we simply divide each tape into 9 blocks and perform a swap for each of the tapes separately. That is: for a $k$-tape Turing machine $M$, we will recursively construct a circuit $C_T$ which, given $k$ $t$-cell configurations with the heads somewhere in the middle third of their tapes, simulates $M$ for $t/3$ steps and outputs the $k$ final $t$-cell configurations. We will repeat the following 3 times, advancing the simulation by $t/9$ steps each time:
\begin{enumerate}
    \item Subdivide the $k$ tapes into 9 equally-sized blocks each.
    \item Identify the indices $i_j \in \set{2, 3, \dots, 8}$, $j \in [k]$ such that head $j$ is in block $i_j$.
    \item For each $j \in [k]$, on tape $j$, swap the blocks $i_j - 1, i_j, i_j +1$ with blocks $1, 2, 3$.
    \item Apply $C_{t/3}$ to blocks $1, 2, 3$ of each of the $k$ tapes (simulating $M$ on that subsection of the tape for $T/9$ steps)
    \item For each $j \in [k]$, on tape $j$, swap the blocks $i_j - 1, i_j, i_j +1$ with blocks $1, 2, 3$.
\end{enumerate}

\end{document}